\documentclass[runningheads]{llncs}
\usepackage[T1]{fontenc}
\usepackage{graphicx,verbatim}
\usepackage{booktabs}
\usepackage{subcaption}
\usepackage{amsmath}
\begin{document}
\title{Real-time, Directionality Aware 3D Ultrasound Reconstruction and Re-Slicing}
\titlerunning{DARE: Directionality-Aware Ultrasound Reslicing}
%
\author{Tobias Jaeggi\inst{1,2} \and
David Black\inst{1} \and
Septimiu Salcudean\inst{1}}
\authorrunning{T. Jaeggi et al.}
%
\institute{University of British Columbia, Vancouver, Canada\\\email{\{dgblack,tims\}@ece.ubc.ca} \and
ETH Zurich, Zurich, Switzerland
}


\maketitle              
\begin{abstract}
Tele-ultrasound through teleoperation allows experts to perform examinations remotely in communities, but limited connectivity can lead to communication delays that reduce usability and diagnostic performance.
Visual-haptic model mediated teleoperation reslices a pre-acquired ultrasound volume in real time to provide an accurate, delay-independent preview image for the sonographer. This enables fast and robust exploration before using the live image for fine tuning. However, existing reslicing techniques do not account for the directional nature of ultrasound - the fact that a structure looks different when imaged from different directions.
This paper presents Directionality-Aware Reslicing (DARE), an ultrasound volume reconstruction and reslicing framework that takes directionality into account. The presented GPU-accelerated algorithm allows real-time reslicing from arbitrary viewpoints to generate accurate preview images. The method is evaluated quantitatively through image similarity metrics and qualitatively through a user study, and significantly outperforms existing reslicing methods in image similarity and realism compared to a ground truth. This can improve the effectiveness and robustness of tele-ultrasound in low-resource areas.

\end{abstract}
\section{Introduction} Ultrasound (US) is a widely used medical imaging technique valued for its portability, low cost, non-invasive nature, and lack of ionizing radiation. However, access to ultrasound imaging remains limited in many remote and rural regions due to shortages of trained personnel. Geographical isolation is a central barrier to US imaging for remote communities, with patients often relying on infrequent visits from itinerant sonographers or long distance travel to obtain imaging~\cite{adams2021access}. 

Tele-ultrasound (Tele-US) aims to improve access by enabling medical experts to remotely guide and perform US exams, but existing approaches face limitations. Video teleguidance systems are simple and inexpensive but inefficient without trained personnel on the patient side~\cite{black2023humanAnon}. Teleoperated robotic or mixed reality tele-US systems enable precise remote control of probe positioning~\cite{jiang2023robotic,black2023humanAnon}. In such systems, a physician teleoperates the exam on a haptic device while viewing the live-streamed US images. The physician thus relies primarily on visual and haptic feedback to perform the exam.

Despite recent advances, tele-US systems remain sensitive to latency, especially in remote communities. Delays of around 150 ms in haptic feedback decrease task performance \cite{kaber2011human}. These effects limit diagnostic accuracy and system usability, but can be reduced using model-mediated teleoperation (MMT), in which the expert interacts haptically with a geometric model of the remote environment \cite{willaert2012stability}. However, visual feedback delays as small as 69 ms have been shown to disrupt haptic manipulation tasks \cite{jay2005delayed}. Latency in ultrasound video transmission can prevent successful identification of imaging targets and increase operator stress \cite{black2025vhmmtAnon,adams2020,masuda2002}. To combat this, in visual-haptic MMT (VHMMT), a 3D ultrasound volume can be constructed either from a pre-acquired sweep of the region of interest performed by the local follower or robot, or from images acquired as the scan progresses. This volume can subsequently be resliced to generate instant 2D ultrasound previews from arbitrary viewpoints locally on the expert console, based on the pose of their haptic device \cite{black2025vhmmtAnon}. This gives a low-latency visual preview that approximates the live ultrasound scan, allowing for fast localization of anatomical structures before using the live image for fine-tuning.

A variety of ultrasound volume reconstruction approaches have been proposed. Pixel-based approaches, such as those in the Public software Library for Ultrasound (PLUS), project 2D US images into a 3D voxel grid and average the intensities of overlapping samples \cite{lasso2014plus}. While computationally efficient and compatible with real time freehand acquisition, sparse data coverage results in gaps requiring compounding or hole-filling techniques \cite{solberg2007freehand}. Function-based methods instead estimate a continuous mathematical function that is fit to model the 3D structure and evaluated on a regular voxel grid, enabling higher-quality reconstructions and improved handling of sparse data. However, it requires solving large optimization problems and operating on substantial data windows, making it computationally expensive and unsuitable for real-time applications \cite{solberg2007freehand}. 

Both approaches treat ultrasound intensity as a scalar quantity and ignore the directional dependency in image formation, despite the strong influence of beam angle on image appearance due to orientation-dependent tissue reflectivity \cite{connolly2001beam}.
Computational sonography was introduced to model the directional dependence by storing orientation-dependent information at each voxel \cite{hennersperger2015computational}. However, its reliance on well-conditioned least-squares estimation limits robustness in sparsely sampled regions and in cases with limited variation in probe orientation. It is also too computationally complex to run in real time. Similarly, neural representations of ultrasound afford good quality reconstructions but require the model to be trained on the new data, making real-time use impractical \cite{liu2026geometry}.
In this work, we propose Directionality Aware Reslicing (DARE), which combines real-time 3D ultrasound volume reconstruction with direction-dependent reslicing. The proposed method is computationally efficient and designed for GPU-based parallel execution, enabling interactive, low-latency reslicing that yields more accurate ultrasound images suitable for tele-ultrasound applications. The main contributions of this paper are:

\begin{itemize}
    \item A directionality-aware volume reconstruction and reslicing framework that preserves probe orientation information and enables more realistic reslicing from arbitrary viewpoint directions 
    \item A real-time implementation of this algorithm on a standard laptop GPU
    \item The evaluation and comparison of the proposed method to a benchmark approach from PLUS
\end{itemize}

\section{Methods}
\subsection{DARE Algorithm}
\subsubsection{Volume reconstruction}
\begin{figure}[h]
    \centering
    \includegraphics[width=0.85\linewidth]{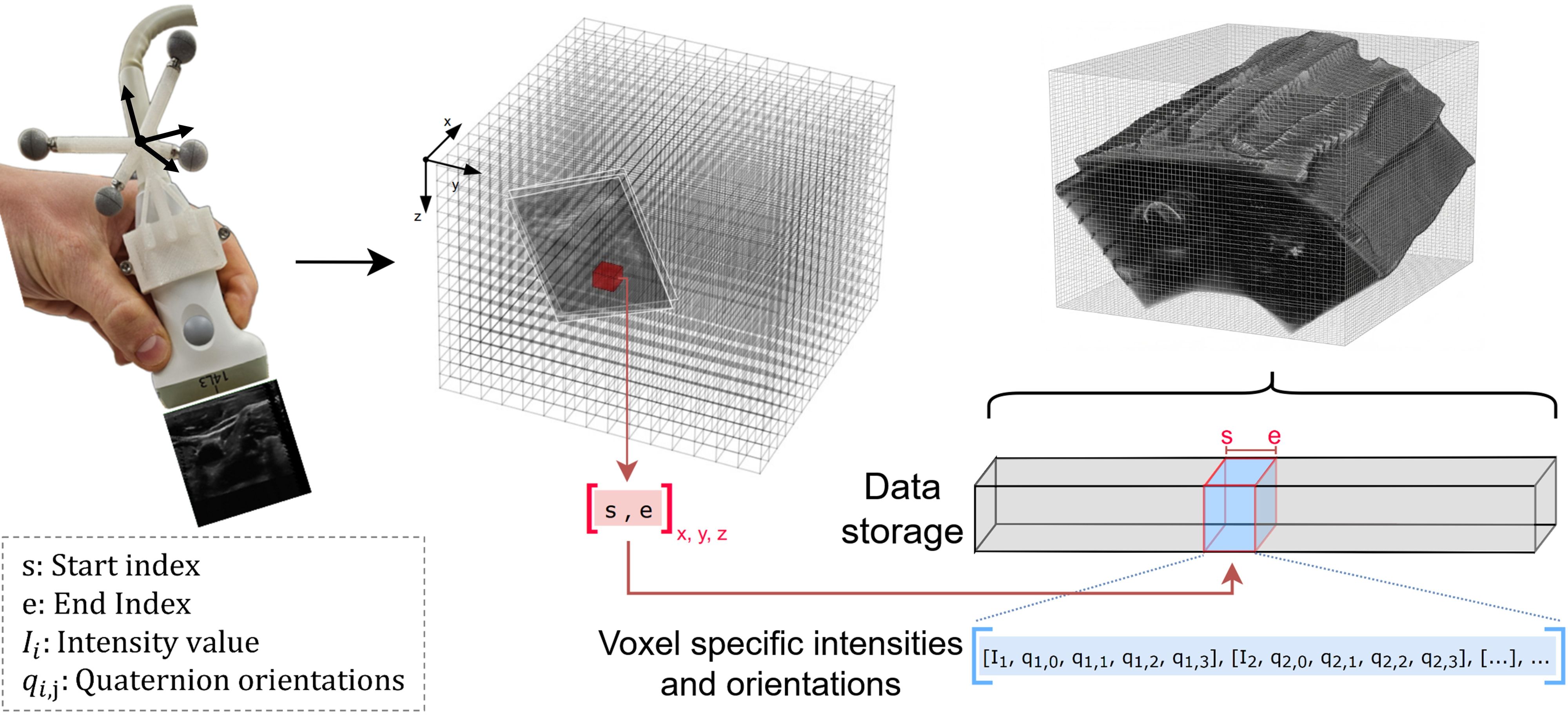}
    \caption{Schematic of the reconstruction process showing the transducer setup including IR markers and the storage approach using a flat structure.}
    \label{fig:reconstruction}
\end{figure}

A parallel process records the live ultrasound image stream and tracks the transducer's position and orientation (pose). The data is stored together with synchronized timestamps to ensure temporal alignment between image frames and spatial tracking data.
After data acquisition, the recorded data is processed through a GPU accelerated algorithm for both reconstruction and reslicing, using CUDA. A 3D bounding box is calculated from the collected transducer positions. Images are transformed to their respective poses relative to the tracking setup and voxelized into a 3D grid. Each grid location points to an index in a flat storage structure that contains the intensity values at that voxel together with the associated acquisition orientations as quaternions. The process is illustrated in Fig. \ref{fig:reconstruction}.

\subsubsection{Reslicing}
During reslicing, a virtual image plane is defined and a 2D ultrasound image is generated by sampling the reconstructed 3D model at the virtual plane's intersection points. For each reslice pixel, candidate samples are gathered from a local cubic neighborhood defined by a fixed interpolation distance.

\begin{figure}[t]
    \centering
    \includegraphics[width=0.85\linewidth]{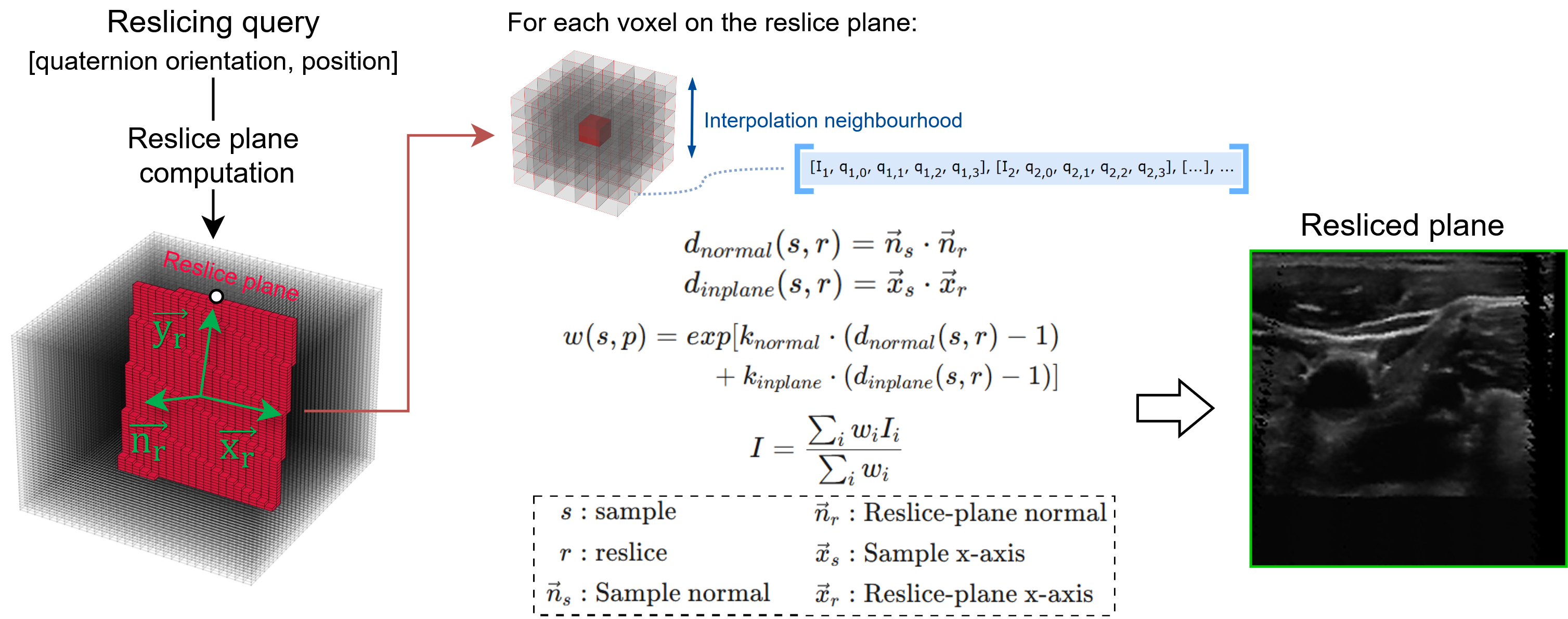}
    \caption{Schematic of the DARE reslicing process, showing plane computation, neighborhood thresholding and exponential weighting by distance and rotation to compute each pixel value.}
    \label{fig:reslicing}
\end{figure}

To incorporate directionality awareness, the relative orientation between each sample's acquisition direction and the reslice plane is evaluated. For each sample $i$, the alignment is quantified using two dot products:
\begin{equation}
d_{\text{normal}}^{(i)} = \vec{n}_{s,i} \cdot \vec{n}_r, \quad d_{\text{inplane}}^{(i)} = |\vec{x}_{s,i} \cdot \vec{x}_r|
\end{equation}
where $\vec{n}_{s,i}$ and $\vec{x}_{s,i}$ denote the sample's acquisition normal and in-plane direction (x-axis) and $\vec{n_r}$, $\vec{x}_r$ denote the corresponding reslice plane vectors (Fig.~\ref{fig:reslicing}).

Samples whose misalignment exceeds predefined angular thresholds are discarded. The remaining samples are combined using an exponentially weighted interpolation:
\begin{equation}
    w_i = \exp\left[k_{\text{normal}}(d_{\text{normal}}^{(i)} - 1) + k_{\text{inplane}}(d_{\text{inplane}}^{(i)} - 1)\right]
\end{equation}
where $k_{normal}$ and $k_{inplane}$ control the sensitivity to orientational deviation. In this work, empirically derived values of $k_{normal}=10$ and $k_{inplane}=5$ were used, with a normal alignment threshold of $25^\circ$ and an in-plane threshold of $15^\circ$. 

The final intensity at each reslice pixel is the exponentially weighted average of the contributing samples. If no valid samples are available, the pixel is left unassigned. This approach suppresses misaligned samples and thereby ensures that only relevant data contributes to the resliced image.

\subsection{Experiments}
To test the approach, ultrasound images were acquired using a BK3500 system with a 14L3 linear array transducer. The probe, mounted with a custom infrared marker array, was tracked using an NDI Polaris Spectra tracking system. To calibrate the optical tracking to ultrasound transformation matrix, the n-wire probe calibration algorithm available in PLUS fCal with the fCal-2.1 phantom \cite{lasso2014plus} was used.

Evaluation was performed using custom semi-cylindrical phantoms, made of gelatin and featuring water-filled plastic tubing to simulate blood vessels, designed to expose internal structures under a wide angular range of probe orientations. Multiple datasets were collected from these phantoms to enable controlled assessments of angular consistency in the resliced images. Additionally, a freehand human dataset was acquired from one of the authors to assess performance on human anatomy. Informed consent was given according to the University of British Columbia Clinical Research Ethics Board, application H23-02587. Each dataset consisted of an initial sweep of the target for volume reconstruction immediately followed by a secondary, different sweep of the same region to obtain target poses for reslicing and corresponding ground-truth reference images.

The proposed DARE method was compared to the PLUS framework, a widely used open source baseline for freehand 3D US reconstruction and reslicing. The same calibration matrix and voxel grid were used for both methods. PLUS used mean compounding with hole-filling at 0.125 mm voxel size, linear interpolation and full optimization. A quantitative analysis was performed on the 552 samples of the human dataset using normalized cross-correlation (NCC) and structural similarity index measure (SSIM) between resliced images and their ground-truth counterparts. Reslice speed was also evaluated by measuring the time required to generate a 2D image out of the reconstructed 3D volume from query to final image output. The evaluation was performed on 100 samples of the human dataset with a DARE reconstruction voxel size of 0.125 mm. 

To assess perceived image quality and diagnostic usability, a two-part user study with three clinical (a radiologist and two sonographers) and five non-clinical (ultrasound researchers) imaging experts was conducted. In the first part, participants compared 17 DARE and PLUS reslices to ground-truth images, assessing image similarity (non-clinical experts) or clinical usefulness (clinical experts). The participants were blinded to the type of reslicing to avoid bias. In the second part, clinical experts selected their preferred interpolation radius (0.125 mm or 0.25 mm) from 9 samples for clinical interpretation. 

\section{Results}
\begin{figure}[h]
    \centering
    \includegraphics[width=0.95\linewidth]{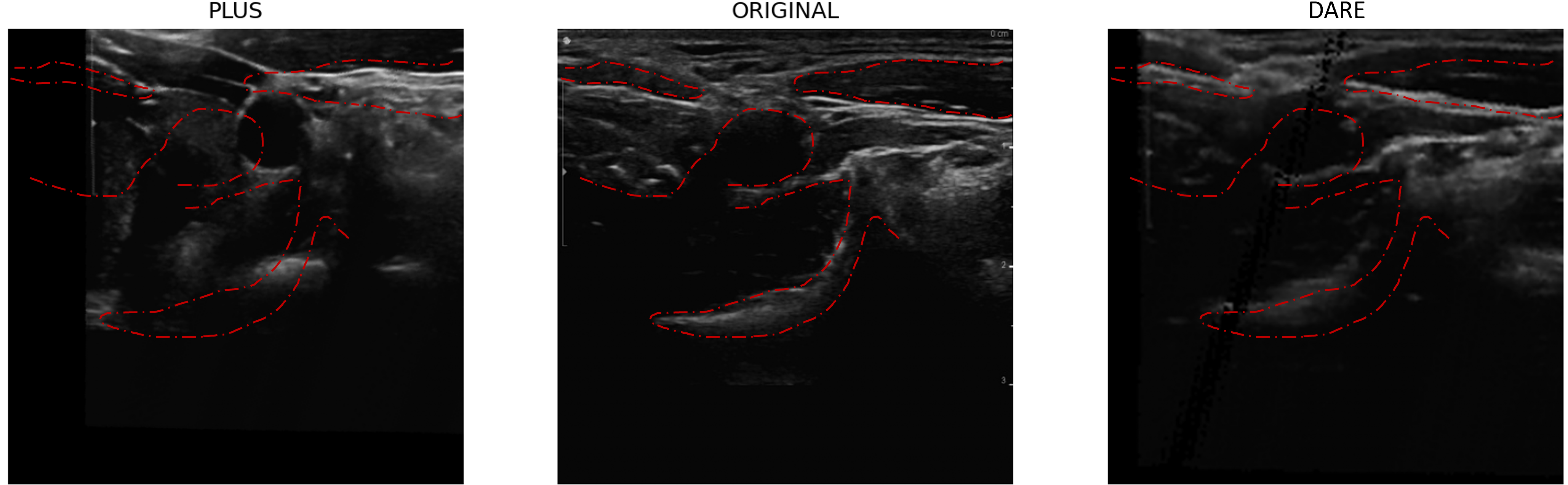}
    \caption{Comparison between PLUS and DARE results on the human scan. DARE shows closer resemblance to the ground-truth than PLUS, where features from other view directions are included. Red lines highlight corresponding features.}
    \label{fig:accuracy}
\end{figure}

\begin{figure}[h]
    \centering
    \includegraphics[width=0.95\linewidth]{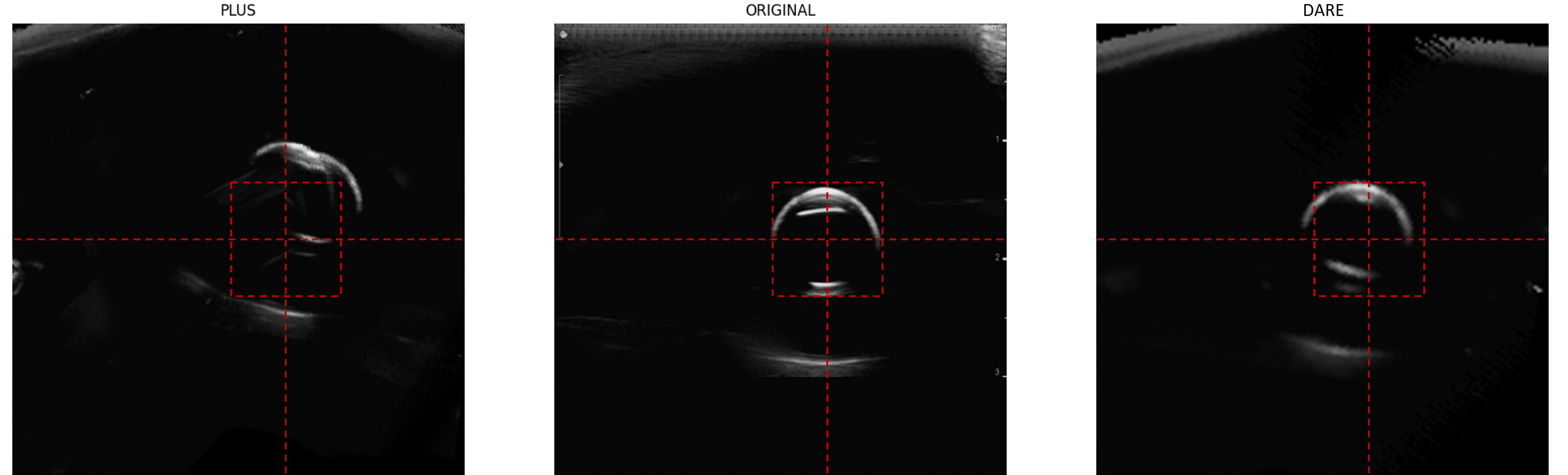}
    \caption{Spatial accuracy comparison of PLUS and DARE versus ground-truth. PLUS shows misalignment and duplicated vessels, while DARE preserves correct features.}
    \label{fig:positionalaccuracy}
\end{figure}

A qualitative visual inspection indicates that DARE produces results that more closely resemble the ground-truth than PLUS. Across both phantom and human datasets, DARE reslices exhibit improved feature depiction and spatial alignment and fewer visible artifacts. This improvement is attributed to the combination of exponentially weighted interpolation and directional thresholding, effectively suppressing contributions from misaligned samples, thus preserving structural coherence and visual clarity (Fig. \ref{fig:accuracy}). Compared to PLUS, which accumulates samples without directional awareness, DARE consistently achieves improved spatial accuracy. As shown in Fig. \ref{fig:positionalaccuracy}, structures imaged from multiple probe orientations appear duplicated and spatially shifted in PLUS reconstruction whereas DARE filters out directionally misaligned samples yielding a single, sharp representation and preserving angular consistency and positional accuracy.

Quantitative image similarity analysis further supports these observations. Compared to ground-truth, DARE reslices achieve higher normalized cross-correlation (NCC) and structural similarity index measure (SSIM) values than PLUS. The median SSIM increases from $0.430$ (IQR: $0.390-0.486$) for PLUS to $0.465$ (IQR: $0.397-0.510$) for DARE, while the median NCC increases from $0.215$ (IQR: $0.164-0.277$) for PLUS to $0.376$ (IQR: $0.323-0.448$) for DARE (paired Wilcoxon signed-rank test, $p<0.001$ for SSIM and NCC). This is despite both metrics penalizing small shifts in the image that are visually and clinically irrelevant. This is clear in Fig. \ref{fig:accuracy}, where the DARE image closely resembles the original, just slightly shifted, whereas the PLUS image contains erroneous structures. The shift causes artificially low similarity scores.

For this reason, the user study, summarized in Table \ref{tab:user_study}, was carried out to assess subjective image quality. All participants preferred the DARE reslices, selecting them most often as the images more similar to ground-truth and more clinically useful. In the second part of the study, a strong consensus emerged among clinical experts favoring an interpolation radius of 0.125 mm, which yields sharper images, over more complete but mildly blurred reslices (0.25 mm). In written feedback, experts emphasized that sharper, minimally processed images better preserve fine anatomical detail for diagnostic assessment.

\begin{table}
\centering
\caption{User study results of non-clinical experts evaluating image similarity and clinical experts evaluating clinical usefulness between DARE and PLUS compared to ground-truth (N: Vote count).}
\label{tab:user_study}
\centering
\scriptsize
\begin{tabular}{lcccc}
    \toprule
    & \multicolumn{2}{c}{\textbf{Non-Clinical Experts}} 
    & \multicolumn{2}{c}{\textbf{Clinical Experts}} \\
    \cmidrule(lr){2-3} \cmidrule(lr){4-5}
    \textbf{Algorithm} 
    & \textbf{N} & \textbf{(\%)} 
    & \textbf{N} & \textbf{(\%)} \\
    \midrule
    PLUS    & 18 & 21\% & 10 & 20\% \\
    DARE    & 61 & 72\% & 28 & 55\% \\
    Both    & 2  & 2\%  & 4  & 8\%  \\
    Neither & 4  & 5\%  & 9  & 18\% \\
    \bottomrule
\end{tabular}
\end{table}

For an interpolation radius of 0.125mm ($3\times3\times3$ voxel neighborhood), the median reslice time was $50\pm7$ ms. Increasing the radius to 0.25 mm and 0.5 mm resulted in median reslice times of $100\pm5$ ms and $300\pm35$ ms, consistent with the cubic growth of the interpolation volume. These results indicate that DARE supports real-time reslicing at the chosen interpolation radius. 

\section{Discussion and Conclusion}
This paper has shown that by incorporating probe directionality, DARE enables low-latency, direction-aware reslicing that preserves spatial consistency and viewpoint-dependent image characteristics. Quantitative image similarity metrics and qualitative evaluations by medical experts show that DARE produces resliced images that more closely resemble ground-truth reference images than conventional approaches. In particular, DARE showed improved robustness to variations in probe orientation and reduced reconstruction artifacts. As a result, VHMMT with DARE has the potential to improve access to ultrasound diagnostics in remote communities and reduce the need for patient transport. However, some limitations remain for further research.

First, the method was compared only to PLUS, not Computational Sonography or neural representations. This is because neither method can operate in real time, making them impractical for tele-ultrasound. However, future work should compare the reconstruction performance explicitly nonetheless. 

Though the median performance of DARE was superior, in some images PLUS outperformed DARE on individual similarity metrics. These cases can be attributed to the sensitivity of similarity metrics to spatial misalignment, as mentioned in the Results. Differences in voxel assignment and interpolation may introduce small spatial shifts on the order of one pixel, which can reduce similarity scores even when structural coherence and perceptual image quality are preserved. In addition, directional filtering and thresholding during DARE reslicing can exclude samples near the orientational thresholds. In sparsely sampled regions, this may result in areas with no contributing samples. In contrast, PLUS aggregates information from all views and applies a hole-filling step, assigning values even in sparsely sampled regions. While this can produce higher similarity scores, it may introduce structural inaccuracies not captured by these metrics. This highlights the limitations of image similarity metrics in capturing quality and motivates the use of qualitative evaluation and expert assessment. 

Prior work reports that latencies above 69 ms degrade operator performance \cite{jay2005delayed}. Expert feedback from the user study indicates that the 0.125 mm interpolation radius provides the most favorable visual result for which the measured latency, 50 ms, remains well below this threshold. The achieved performance is enabled by GPU acceleration. While initial CPU-based implementations required several seconds per reslice, with more complex interpolation exceeding one minute, the GPU implementation reduced computation time by over two orders of magnitude, making real-time visualization feasible. The requirement for a GPU does not limit accessibility in low-resource areas since the reslicing is performed on the physician-side, not in the remote site.

In the user assessment, medical experts selected the option “Neither” more frequently than non-clinical experts. This indicates that, despite the overall preference for DARE, the resliced images do not yet consistently meet clinical expectations. However, the proposed method is not intended to replace live ultrasound imaging. Instead, it is designed as a low-latency visual preview that assists the expert in rapidly finding an anatomical target location before switching to live imaging for diagnosis. In this context, small artifacts are less critical than preserving correct spatial alignment and overall anatomical coherence.

Moreover, in practical ultrasound examinations, sonographers routinely apply a wide range of forces to the transducer to acquire diagnostically useful images. Pressure-induced tissue deformation directly affects the spatial position of imaged structures, posing significant challenges for volume reconstruction. Since deformation along the probe force direction is not accounted for in the current setting, pressure variations may lead to overlapping features and spatial inconsistencies. This highlights the need for future extensions to the reconstruction approach to account for probe-tissue interaction effects.

To capture the dynamics of a real tele-US scenario, factors such as patient movement, interactive probe control, and closed-loop feedback must be included. The present work reconstructs an ultrasound volume from a pre-acquired sweep that is used as a static model for evaluation. This design choice enables controlled analysis of reslicing performance and directionality effects. However, it does not yet constitute a fully integrated clinical workflow. The current framework should be extended in future work to enable incremental integration of newly acquired data during a scan, and to support increasing dataset sizes despite constraints on GPU memory. This would enable the local visual model to evolve over time and give the expert more complete and higher quality resliced images. 

Finally, evaluating the proposed approach in closed-loop teleoperation scenarios by using either robotic or human-teleoperated probe control on real patients with artificially introduced communication delays would provide valuable insights into the practical benefits of the VHMMT framework with DARE in tele-ultrasound applications. This work can enable more robust, faster tele-ultrasound in the future to improve healthcare access in rural areas.

\begin{credits}
\subsubsection{\discintname}
The authors have no competing interests to declare that are relevant to the content of this article.
\end{credits}

\bibliographystyle{splncs04}
\bibliography{refs}
\end{document}